\documentclass[12pt]{article} \hoffset=-15mm \voffset=-10mm
\usepackage{graphicx,amsmath,amssymb,epsf} \usepackage{latexsym,bm,
  slashed} \usepackage{xcolor} \definecolor{dark}{rgb}{0.10,0.2,0.3}
\definecolor{magenta}{rgb}{0.7,0.1,0.3}
\definecolor{purpure}{rgb}{0.5,0.15,0.3}
\usepackage[font=small,format=plain,labelfont=bf,up,textfont=it,up]{caption}
\usepackage{hyperref, cite} \hypersetup{colorlinks, citecolor=blue,
  filecolor=blue, linkcolor=magenta,
  urlcolor=purpure,hyperfootnotes=true,pdftex}

\def\be{\begin{equation}}
\def\ee{\end{equation}}
\def\bea{\begin{eqnarray}}
\def\eea{\end{eqnarray}}

\def\ubar{\bar{u}}

\newcommand{\slv}{\raise.15ex\hbox{$/$}\kern-.53em\hbox{$v$}}
\newcommand{\sln}{\raise.15ex\hbox{$/$}\kern-.53em\hbox{$n$}}
\newcommand{\slnbar}{\raise.15ex\hbox{$/$}\kern-.53em\hbox{$\bar{n}$}}
\newcommand{\slF}{\raise.15ex\hbox{$/$}\kern-.53em\hbox{$F$}}
\newcommand{\sll}{\raise.15ex\hbox{$/$}\kern-.40em\hbox{$l$}}
\newcommand{\slh}{\raise.15ex\hbox{$/$}\kern-.40em\hbox{$h$}}
\newcommand{\slP}{\raise.15ex\hbox{$/$}\kern-.53em\hbox{$P$}}
\newcommand{\slp}{\raise.15ex\hbox{$/$}\kern-.53em\hbox{$p$}}
\newcommand{\slq}{\raise.15ex\hbox{$/$}\kern-.53em\hbox{$q$}}
\newcommand{\slR}{\raise.15ex\hbox{$/$}\kern-.53em\hbox{$R$}}
\newcommand{\slz}{\raise.15ex\hbox{$/$}\kern-.53em\hbox{$Z$}}
\newcommand{\slzbar}{\raise.15ex\hbox{$/$}\kern-.53em\hbox{$\bar{Z}$}}
\newcommand{\slQ}{\raise.15ex\hbox{$/$}\kern-.53em\hbox{$Q$}}
\newcommand{\slK}{\raise.15ex\hbox{$/$}\kern-.53em\hbox{$K$}}
\newcommand{\slk}{\raise.15ex\hbox{$/$}\kern-.53em\hbox{$k$}}
\newcommand{\slkbar}{\raise.15ex\hbox{$/$}\kern-.53em\hbox{$\bar{k}$}}
\newcommand{\slkone}{\raise.15ex\hbox{$/$}\kern-.53em\hbox{$k_1$}}
\newcommand{\slpone}{\raise.15ex\hbox{$/$}\kern-.53em\hbox{$p_1$}}
\newcommand{\slpbarone}{\raise.15ex\hbox{$/$}\kern-.53em\hbox{$\bar{p}_1$}}
\newcommand{\slptwo}{\raise.15ex\hbox{$/$}\kern-.53em\hbox{$p_2$}}
\newcommand{\slpbartwo}{\raise.15ex\hbox{$/$}\kern-.53em\hbox{$\bar{p}_2$}}
\newcommand{\slqone}{\raise.15ex\hbox{$/$}\kern-.53em\hbox{$q_1$}}
\newcommand{\slD}{\raise.15ex\hbox{$/$}\kern-.53em\hbox{$\!D$}}
\newcommand{\slC}{\raise.15ex\hbox{$/$}\kern-.53em\hbox{$C$}}
\newcommand{\slA}{\raise.15ex\hbox{$/$}\kern-.73em\hbox{$A$}}
\newcommand{\slSigma}{\raise.15ex\hbox{$/$}\kern-.53em\hbox{$\Sigma$}}
\newcommand{\slpartial}{\raise.15ex\hbox{$/$}\kern-.53em\hbox{$\partial$}}
\newcommand{\slcalP}{\raise.15ex\hbox{$/$}\kern-.63em\hbox{$\cal P$}}
\newcommand{\sleps}{\raise.15ex\hbox{$/$}\kern-.53em\hbox{$\epsilon$}}
\newcommand{\slepsbar}{\raise.15ex\hbox{$/$}\kern-.53em\hbox{$\overline{\epsilon}$}}
\newcommand{\slepsstar}{\raise.15ex\hbox{$/$}\kern-.53em\hbox{$\epsilon$}^\star}

\newcommand{\nn}{\nonumber\\}

 \title{\bf
  \Large \bf \Large Elastic scattering of a quark from a color field: longitudinal momentum exchange} \author{ Jamal~Jalilian-Marian$^{1,2,3}$
  \bigskip \\
  {\normalsize $^1$Department of Natural Sciences, Baruch College,
    CUNY,}
  \\
  {\normalsize 17 Lexington Avenue, New York, NY 10010, USA}\\
  {\normalsize $^2$CUNY Graduate Center, 365 Fifth Avenue, New York, NY 10016, USA}\\
  {\normalsize $^3$Centre de Physique Th\'eorique, \'Ecole
    Polytechnique, CNRS,}
  \\
  {\normalsize Universit\'e Paris-Saclay, 91128 Palaiseau, France} }

\begin{document}

\maketitle
\begin{abstract}

Perturbative QCD in the small Bjorken $x$ limit can be formulated as an 
effective theory known as the Color Glass Condensate (CGC) formalism. The 
CGC formalism takes into account the dynamics of large gluon densities 
at small $x$ and has been successfully applied to Deep Inelastic 
Scattering (DIS) and particle production in high energy hadronic and 
nuclear collisions in the small $x$ kinematic region. The effective 
degrees of of freedom in CGC are Wilson lines which enter in the 
effective quark (and gluon) propagators and re-sum multiple soft 
scatterings from the small $x$ gluon field of the target. It is 
however known that the CGC effective theory breaks down when one probes 
the moderately large $x$ (high $p_t$) kinematics where collinear factorization 
and DGLAP evolution of parton distribution functions should be 
the right framework. Here we propose a general framework which 
may allow one to eventually unify the two approaches and to 
calculate pQCD cross sections in both small and large Bjorken $x$ 
regions. We take the first step towards this goal by deriving an 
expression for the quark 
propagator in a background field which includes scatterings from 
both small and large $x$ modes of the gluon field of the target. 
We describe how this quark propagator can be used to calculate 
QCD structure functions $F_2$ and $F_L$ at all $x$ and thus 
generalize the dipole model of DIS. We outline this approach can 
also be used to extend the so-called hybrid approach to 
particle production in the forward rapidity region of high energy 
hadronic and nuclear collisions to all $x$ and $p_t$ regions and 
speculate on how one may apply the same techniques to extend 
the McLerran-Venugopalan effective action used in high energy 
heavy ion collisions to include high $p_t$ physics.

\end{abstract}

\section{Introduction}

Parton (gluon) distribution functions are known to 
grow very fast at small Bjorken $x$ which would lead to a violation 
of the Froissart bound on growth of physical cross sections with energy. 
Gluon saturation was proposed~\cite{glr,mq} as a dynamical mechanism which can 
tame this growth and restore perturbative unitarity. The Color Glass 
Condensate (CGC) formalism~\cite{mv} is an effective action approach 
to gluon saturation in QCD that includes coherent multiple scatterings from 
the gluon fields of the target, which are essential when gluon density 
of the target is large, i.e. at small $x$. 
It also includes high energy effects by re-summing large logs of $1/x$ via 
the JIMWLK/BK evolution equation~\cite{jimwlk,bk} and as such it differs from 
the collinear factorization approach to particle production in pQCD where 
parton distribution functions evolve according to the DGLAP evolution 
equation~\cite{DGLAP}.

The CGC formalism has been developed significantly since its inception 
both in terms of precision (higher order corrections) as well as the range of 
physical processes considered (see \cite{review-cgc} for references). Perhaps 
the most important aspect of the CGC 
formalism is the emergence of a dynamically generated and potentially large 
scale, the saturation scale $Q(x, A, b_t)$, which allows one to understand 
a wide range of phenomena in high energy QCD by making controlled approximations 
and quantitatively reliable calculations. 

Despite many successes of the CGC formalism, its kinematic domain of 
applicability remains limited; at best it may be applied to processes where the 
cross section is dominated by small $x$ gluons (as a rough guide one can take  
small $x$ to mean $x < 0.01$). Recalling that in particle production in high 
energy hadronic/nuclear collisions $x$ and $p_t$ are kinematically correlated, 
the dominant contribution to high $p_t$ processes is from the not so small $x$ 
region. Therefore high $p_t$ particle production is currently not computable 
in CGC formalism where one must have $\log 1/x \gg \log p_t^2$. Furthermore it's 
been realized quite recently that CGC 
calculations of higher order corrections to particle production cross sections 
are not stable and can become negative~\cite{cgc-negative} at transverse 
momenta slightly higher 
than the saturation scale $Q_s (x)$. While there has been 
various remedies proposed to cure this problem~\cite{cgc-negative} in 
case of single inclusive 
hadron production in asymmetric collisions (such as proton-proton and 
proton-nucleus collisions in the forward rapidity region in the hybrid 
approach~\cite{hybrid}), 
the general situation remains unsatisfactory and is likely an 
indication that at high 
$p_t$ important physics is missing. One such deficiency is the physics of 
DGLAP evolution which re-sums potentially large logs of $p_t^2$ in parton 
distribution functions in the context of collinearly factorized cross sections 
in pQCD in the leading twist approximation. This is the dominant effect at high 
$p_t$ where $\log p_t^2 \gg \log 1/x$ and is not included in the CGC formalism. 
The two approaches coincide when $\log p_t^2 \sim \log 1/x$, known 
as the double log limit. 

Ideally one would like to have a unified formalism for particle production
in QCD which reduces to collinear factorization and DGLAP evolution equation 
at intermediate/large $x$ and/or $p_t$ and to CGC and JIMWLK equation at 
small $x, p_t$ (see \cite{bal} for instance). In collinear factorization/DGLAP 
formalism quarks and gluons and their distribution functions are the degrees 
of freedom used while in the small $x$ limit it is most efficient to use 
(fundamental or adjoint) Wilson lines, path ordered exponentials of gluon 
field which re-sum coherent multiple scatterings of a quark or gluon on a 
soft color field. This Wilson line
is very closely connected to the quark/gluon propagator in a background color
field~\cite{fg}. One may then 
naturally ask, what would be the effective degrees of freedom in a unified 
approach? In other words, what could be the building blocks of a physical cross
section in a more general approach which has both small $x$ and large $x$
combined?

In the small $x$ limit one makes a drastic approximation by taking
the $x\rightarrow 0$ limit, for example, in parton splitting functions whereas
in DGLAP approach the full splitting function is kept. Clearly to have any
hope of a unified approach, one must relax the $x\rightarrow 0$ limit. As
$x$ is the ratio of longitudinal momenta or energies of the partons in the
parent proton/nucleus, this means that one must consider scattering of a
projectile parton not only from the small $x$ modes of the target 
proton/nucleus but
also from the more energetic modes at large $x$.
Furthermore, DGLAP is a leading twist formalism so that multiple exchanges 
that involve extra powers of the hard scale are dropped. Keeping these two
points in mind, we propose a new approach which may enable us to 
eventually "unify" the two formalisms such that one recovers the CGC 
formalism and JIMWLK evolution equation at small $x$ and pQCD collinear 
factorization and DGLAP evolution equation at large $x$. 

Toward this end we consider here scattering of a quark from a background color 
field which is, unlike in the small $x$ limit, allowed
to have both large and small $x$ modes. We re-sum the multiple scatterings from
the fields with small $x$ modes to all orders while keeping only one scattering 
from large $x$ modes of the field. Therefore the scattered quark can in general
carry any energy and transverse momentum and be deflected by large angles, unlike 
the small $x$ limit where the scattering is eikonal (energy of the quark remains 
the same) and limited to small transverse momentum exchanges. From the 
calculated quark 
scattering amplitude we extract the effective quark propagator in this more 
general background field. We show how this new quark propagator can be used to 
generalize the dipole model of the QCD structure functions $F_2, F_L$ 
at small $x$. We expect that the resulting expressions would form the starting 
point for calculating the one-loop correction to the structure functions 
which would then result in a more general evolution equation containing 
JIMWLK evolution at small $x$ and DGLAP evolution at intermediate/large $x$. 
We outline how this new approach may be used to generalize 
the CGC framework to include particle production at high $p_t$.

\section{Multiple scattering at small $x$: eikonal approximation}
\label{sec:eikonal}

In this section we briefly review multiple scattering in the eikonal 
limit and its use in small $x$ physics. There are already excellent 
reviews covering eikonal scattering~\cite{eik-review} so everything 
in this section is already well-known. Therefore here we use the example of 
eikonal scattering to remind the reader of the approximations and 
assumptions made at small $x$, and to set up our formalism. In this
section we closely follow the approach of Casalderrey-Solana and
Salgado in section $4.1$ of \cite{eik-review} (see also section $4$
of the first reference in \cite{fg}).

The light cone coordinates are defined as
\be
x^+ \equiv {t + z \over \sqrt{2}}\,\, , \, \, 
x^- \equiv {t - z \over \sqrt{2}}
\ee
and similarly for momenta and fields. We start
by considering scattering of a high energy quark moving in the positive 
$z$ direction, on a high energy target (proton/nucleus) moving to the left
as shown in Fig. ({\ref{fig:light-cone}).
The quark has momentum $p^\mu = (p^+ \sim \sqrt{s},\, p^- = 0,\, p_t =0)$ 
whereas the target has momentum 
$P^\mu = (P^+ = 0,\, P^- \sim \sqrt{s},\, P_t = 0)$ even though 
what's really meant in the eikonal approximation is that $p^+ \gg p^-, p_t$ 
and likewise for the target momentum. 

\begin{figure}[h]
  \centering
  \includegraphics[width=.4\textwidth]{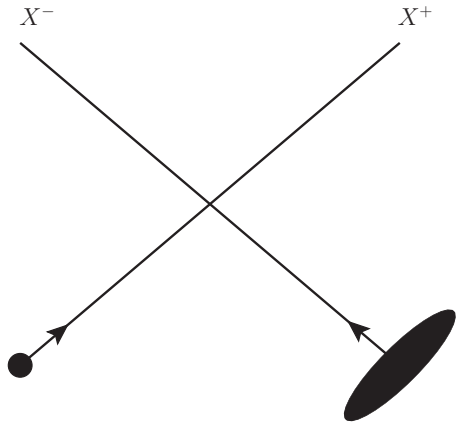}
  \caption{\it Quark scattering on a target nucleus or proton}
  \label{fig:light-cone}
\end{figure}

In the CGC formalism one thinks of the target as a current of color charges which has 
only one large component, $J^\mu_a \simeq \delta^{\mu -} \rho_a$. The color current 
is covariantly conserved and satisfies 
\be
D_\mu \, J^\mu = D_- \, J^- = 0
\label{eq:cur-cons}
\ee
If we now choose to work in the light cone gauge $A^+ = 0$, we get $\partial_- J^- =0$
which means the color current (charge) is independent of the coordinate $x^-$,
\be
J^{\, -} = J^{\, -} (x^+, x_t)
\ee
To find the color fields generated by this color current one can solve the classical
equations of motion 
\be
D_\mu\, F^{\mu \nu} = J^\nu
\ee
in the light cone gauge $A^+ = 0$ and find the classical solution $A^\mu$ in 
terms of the color charge $\rho$. The only non-zero component of the field is
$A^-$ which is independent of the coordinate $x^-$,
\be
A^- = A^- (x^+, x_t)
\ee
so that in momentum space   
\be
A^- = A^- (p^+ \sim 0, p^-, p_t).
\ee 
This is important in eikonal scattering, the target 
field carries almost no plus momentum so that it can not impart any plus 
momentum to the projectile. As a result the plus momentum of the projectile 
is conserved as we will see in detail below. Also related, the small $x$ 
momentum modes of the target (moving in the negative $z$ direction) carry 
negligible $P^-$ momentum and therefore can not impart any sizable minus
component of momentum to the projectile. So basically the only momentum 
exchanged in the eikonal limit is small transverse momenta. 

An equivalent 
way to understand this to consider the coupling of the gauge field $A^\mu$ 
to the fermion current $\bar{u} (p_f)\, \gamma_\mu \, u (p_i)$. In the 
eikonal approximation one takes $p_f \simeq p_i$ and using 
$\bar{u} (p)\, \gamma_\mu \, u (p) \sim p_\mu$ we see that if the projectile
quark has a large $p^+$ it couples only to $A^-$ component of the gauge
field in eikonal scattering. This is a general result; the quark moving
along the $"+"$ direction in its light cone frame will couple only to the
$"-"$ component of the filed in that frame. This is extremely important 
and will be used extensively here.
  
Clearly to include large $x$ modes of the target one needs to 
allow exchange of longitudinal momentum which amounts to 
scattering of a quark not only from the small $x$ modes of the 
target but also from the large $x$ modes carrying a finite 
fraction of the target energy $P^-$. We will consider this in 
the next section. Therefore to study scattering of a quark on a 
target at small $x$ we consider scattering (propagation) of a 
quark in a background field $A^-$ which is independent of 
coordinate $x^-$. Each scattering costs a power of $g$ so that if 
$A^- \sim 1/g$ then $g\, A^- \sim \mathcal{O}\,(1)$ 
and one needs to re-sum all such scatterings. Note that we haven't made 
any assumption about the $x^+$ dependence of the fields which in the 
extreme limit is usually taken to be a delta function due to Lorentz 
contraction of the target. Here we will keep the $x^+$ dependence 
general to allow for an extended target. To proceed it is useful to 
define a light-like vector $n^\mu$ which points in the $x^-$ direction 
so that $n^\mu = (n^+ = 0, n^- = 1, n_t = 0)$ and $n^2 = 0$ and that 
$n\cdot A = 0$ defines our gauge with $\sln = \gamma^+$. Using this 
light-like vector one can also write 
$A^-_a (x^+, x_t)= n^-\, S_a (x^+, x_t)$ so that 
$\slA = \sln\, S = S\, \sln$. So basically the Lorentz 
index of the field $A^-$ is carried by the vector $n^\mu$. 

To start we consider multiple scattering of a quark on the target 
color fields one scattering at a time; momentum of the incoming 
quark is denoted $p$, while $q$ is its outgoing momentum. To make 
the approximations made clear (and so that we don't have to repeat the 
details in the next section) we will show all the details of the 
calculation here. The amplitude for one scattering is shown in 
Fig. (\ref{fig:one-scatt}) 
\begin{figure}[h]
  \centering
  \includegraphics[width=.5\textwidth]{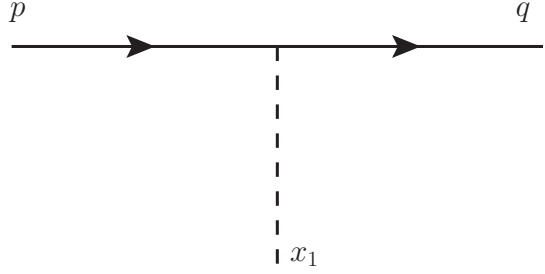}
  \caption{\it Scattering of a quark from a soft field at position 
$x_1^\mu$. The target is not drawn explicitly for brevity.}
  \label{fig:one-scatt}
\end{figure}
where solid line denotes a quark while the dashed line denotes the 
soft gluon field of the target at coordinate $x_1^\mu = (x_1^+, x_{1t})$.
The target is represented by point $x_1$ and not shown for brevity. 
The amplitude is 
\bea
i \mathcal{M}_1\!\!\! &=&\!\!\! (i g) \int d^4 x_1\, e^{i (q - p) x_1}\, \ubar (q)\, 
\left[\sln \, S (x_1)\right]\, u(p)\nn
\!\!\! &=&\!\!\!
(i g)
(2\pi) \delta (p^+ - q^+)\!\! \int d^2 x_{1t}\, d x_1^+\, 
e^{i (q^- - p^-) x_1^+}\, e^{- i (q_t - p_t) x_{1t}}\, 
\ubar (q) 
\left[\sln \, S (x_1^+, x_{1t})\right] u(p)\, . \nn
&&
\eea
Since the field $S$ does not depend on coordinate $x^-$ the integration over $x^-$ is
trivial and gives an overall delta function $2 \pi \delta (p^+ - q^+)$ so that the plus 
momentum of the quark is conserved. Including a second scattering is depicted in 
Fig. (\ref{fig:two-scatt}) and gives
\be
i \mathcal{M}_2 = (i g)^2 \int d^4 x_1\, d^4 x_2 \, 
\int {d^4 p_1 \over (2 \pi)^4}\, 
e^{i (p_1 - p) x_1}\, 
e^{i (q - p_1) x_2}\, 
\ubar (q)\, \left[ \sln \, S (x_2)\, 
{i \slpone \over p_1^2 + i \epsilon}\, \sln \, S (x_1) 
\right]\, u(p)\, .
\nn
\ee

\begin{figure}[h]
  \centering
  \includegraphics[width=.5\textwidth]{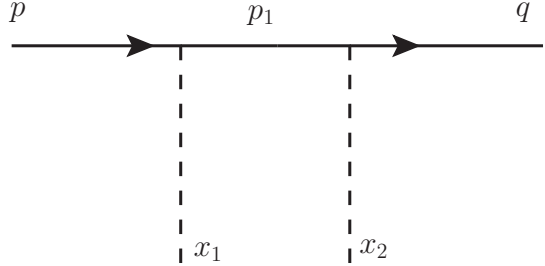}
  \caption{\it scatterings of a quark on two soft fields.}
  \label{fig:two-scatt}
\end{figure}
Once again one can perform the $x_1^-, x_2^-$ integration giving two delta 
functions $2 \pi \delta (p_1^+ - q^+)\, 2 \pi \delta (q^+ - p_1^+)$ one of which can be
used to perform the $p_1^+$ integration setting $p_1^+ = p^+ = q^+$ with an overall 
$2 \pi \delta (p^+ - q^+)$ remaining. We get
\bea
i \mathcal{M}_2 &=& (i g)^2 \, 2 \pi \delta (p^+ - q^+)\, 
\int d^2 x_{1t}\, d x_1^+ d^2 x_{2t}\, d x_2^+ \, 
\int {d^2 p_{1t} \over (2 \pi)^2}\, {d p_{1}^- \over (2 \pi)}\,
e^{i (p_1^- - p^-) x_1^+}\, e^{- i (p_{1t} - p_t) \cdot x_{1t}}\nn
&& e^{i (q^- - p_1^-) x_2^+}\, e^{- i (q_t - p_{1t}) \cdot x_{2t}}\, 
\ubar (q)\, \left[ S (x_2^+, x_{2t})\, \sln \, 
{i \slpone \over p_1^2 + i \epsilon}\, \sln \, S (x_1^+, x_{1t}) 
\right]\, u(p)
\eea
and $p_1^+ = p^+ = q^+$ in the propagator. The next step is to perform the integral
over $p_1^-$. This can be done via contour integration noticing that $\slpone$ in the
numerator is next to $\sln$ and that $\sln \, p_1^- \gamma^+ = p_1^- \, \sln \sln = 0$.
Therefore the integration over $p_1^-$ can be performed by closing the contour below
the real axis and gives
\be
\int {d p_{1}^- \over (2 \pi)} 
{e^{i p_1^- (x_1^+ - x_2^+)} \over 
2 p^+ \left[p_1^- - {p_{1t}^2 - i \epsilon \over 2 p^+}\right]} =
{- i \over 2 p^+} \, \theta (x_2^+ - x_1^+) \, 
e^{i {p_{1t}^2 \over 2 p^+} (x_1^+ - x_2^+)}
\ee
There are two points which we should mention here, first an incoming quark with 
positive $p^+$ forces ordering along the direction of motion $x^+$, enforced 
via the theta function, and second, in the strict eikonal approximation one 
disregards the exponential on the right hand side by using the fact that 
${p_{1t} \over p^+} \ll 1$. It is possible to go beyond this strict eikonal 
limit and keep these exponentials factors (for example see~\cite{eik-review}), 
however we will not do that here. We then have
\bea
i \mathcal{M}_2 &=& (i g)^2 \, (- i ) \, 2 \pi \delta (p^+ - q^+)\, 
\int d^2 x_{1t}\, d x_1^+ d^2 x_{2t}\, d x_2^+ \, \theta (x_2^+ - x_1^+) \,
e^{- i p^- x_1^+}\, e^{i q^- x_2^+}\nn
&& \int {d^2 p_{1t} \over (2 \pi)^2}\,
e^{- i (p_{1t} - p_t) \cdot x_{1t}}\, e^{- i (q_t - p_{1t}) \cdot x_{2t}}\, 
\ubar (q)\, \left[ S (x_2^+, x_{2t})\, \sln \, 
{i \slpone \over 2 p^+} \, \sln \, S (x_1^+, x_{1t}) 
\right]\, u(p)\nn
&&
\eea
There are two further approximations we need to make before we get the final 
result, first, we note that $p^- = {p_t^2 \over 2 p^+} \ll 1$ and 
$q^- = {q_t^2 \over 2 q^+} \ll 1$ and one therefore can drop the exponential 
terms involving $p^-, q^-$ above. More importantly, there is still a $p_{1t}$
term in the intermediate propagator which we also neglect since it is of the form 
${p_{1t} \over p^+} \ll 1$  and can be dropped (it is possible to keep 
these terms, order by order, see~\cite{b-eik} for example). After this last 
step, one can perform the integration over transverse momentum $p_{1t}$ 
which gives
\be
\int {d^2 p_{1t} \over (2 \pi)^2} \, e^{- i p_{1t} \cdot (x_{1t} - x_{2t})}
= (2 \pi)^2 \delta^2 (x_{1t} - x_{2t})
\ee
and enables one to perform one of the transverse coordinate integrations to 
get
\bea
i \mathcal{M}_2 &=& (i g)^2 \, (- i ) ( i ) \, 2 \pi \delta (p^+ - q^+)\,  
 \int \, d x_1^+ \, d x_2^+ \, \theta (x_2^+ - x_1^+) \,
\int d^2 x_{1t}\,
e^{- i (q_t - p_t) \cdot x_{1t}}\nn
&&
\ubar (q)\, \left[ S (x_2^+, x_{1t})\, \sln \, 
{\slpone \over 2 p^+} \, \sln \, S (x_1^+, x_{1t}) 
\right]\, u(p)
\eea
where $\slpone$ is now understood to be $p_1^+ \, \gamma^-$. The last step is
to use $\sln \, {\slpone \over 2 p^+} \, \sln = \sln$ since $p_1^+ = p^+$.
We note that the two small $x$ fields $S$ are at the same transverse coordinate $x_{1t}$
which was made possible due to neglecting terms ${p_{1t} \over p^+}$ in both 
the exponentials and in the intermediate propagator. The second point to be kept in
mind is the ordering of the scatterings which was caused by positivity of $p^+$ of
the incoming quark and that contour integration over the propagator pole puts the
intermediate quark line on shell. 

The approximations and steps needed to extend this procedure to include any number of scatterings from the small $x$ fields $S$ should now be clear; the energetic 
projectile moving in a given direction will couple to the component of the gluon 
field Lorentz-conjugate to that direction, for example $p^+ \, :\, A^-$. One
neglects terms of the order ${p_t \over p^+}$ everywhere and that the $p^+$ 
component of the 
projectile momentum is conserved. The generic diagram for $n$ soft scatterings is
shown in Fig. (\ref{fig:n-scatt}), 
\begin{figure}[h]
  \centering
  \includegraphics[width=1.0\textwidth]{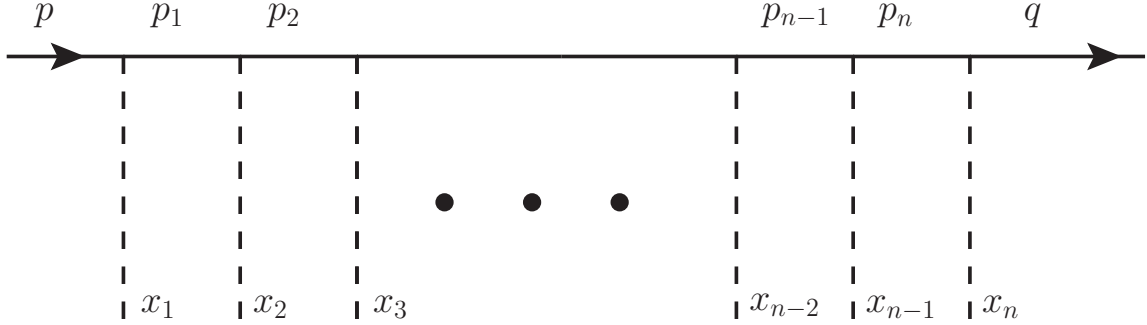}
  \caption{\it N soft scatterings of a quark.}
  \label{fig:n-scatt}
\end{figure}
The amplitude is then
\bea
i \mathcal{M}_n &=& 2 \pi \delta (p^+ - q^+)\,  
\ubar (q)\, \sln\, \int d^2 x_{t}\, e^{- i (q_t - p_t) \cdot x_{t}} 
\nn 
&& 
\bigg\{
(i g)^n \, (- i )^n ( i )^n\, 
\int \, d x_1^+ \, d x_2^+ \, \cdots d x_n^+ \, 
\theta (x_n^+ - x_{n - 1}^+) \, \cdots \theta (x_2^+ - x_1^+)\nn
&&
\left[ S (x_n^+, x_{t})\, S (x_{n - 1}^+, x_{t})\,  
\cdots S (x_2^+, x_{t})  S (x_1^+, x_{t}) 
\right]\bigg\} \, u(p)
\eea
which after summing over all $n$ scatterings 
($i\mathcal{M} = \sum\limits_{n=1}^{\infty} i \mathcal{M}_n$) gives 
the standard result for quark-target scattering amplitude in the 
eikonal limit~\cite{adjjm2},
\be
i \mathcal{M} (p,q) = 2 \pi \delta (p^+ - q^+)\,  
\ubar (q)\, \sln\, \int d^2 x_{t}\, e^{- i (q_t - p_t) \cdot x_{t}} \, 
\left[V (x_t) - 1\right]\, u(p)
\ee
where $V (x_t)$ is the Wilson line in the fundamental representation, 
defined as~\footnote{Here we are being sloppy about path ordering vs 
anti-path ordering of the fields. The Wilson lines appearing in the 
propagators are path ordered, i.e., the soft fields increase in their 
$x^+$ argument from left to right, whereas the Wilson lines in the 
amplitudes are anti-path ordered.
As introducing different symbols for path-ordered vs anti path-ordered 
will result in clutter and because it will not have any bearing on 
our result we will ignore this distinction.}
\be
V (x_t) \equiv \hat{P}\, 
\exp \left\{i g \int_{- \infty}^{+\infty} d x^+ \, S^-_a (x^+, x_t)\, t_a\right\}
\ee
and $S^- (x^+, x_t) = n^-\, S (x^+, x_t)$ is our small $x$ gluon field.  

One can extract an effective quark propagator using the result for 
scattering amplitude. To do so we recall that the incoming 
projectile was a quark with positive $p^+$ which is all we need in 
case of scattering of a physical particle. However, to construct a 
Feynman propagator one also needs to consider the case when a 
particle with $- p^+$ is moving backward (what we have computed so far
would give us the retarded and not the Feynman propagator). This is 
practically trivial to implement since the only change in the 
derivation above is the locations of $p^-$ poles of the intermediate 
propagators which will be above 
the real axis as compared to before where the poles were below the 
real axis. This 
reverses the ordering of projectile propagation in $x^+$ direction. 
It also results in a relative minus sign between the positive and 
negative $p^+$ contributions due to the reversal of the direction 
of the integration contour when the contour is closed above vs 
below the real axis.

Defining (recall that in calculating the amplitude one goes from right to left
in a Feynman diagram while for the propagator we go from left to right)
\be
\label{eq:tau_eik}
\tau_F (p,q) \equiv  2 \pi \delta (p^+ - q^+)\,  
\sln\, \int d^2 x_{t}\, e^{- i (q_t - p_t) \cdot x_{t}} \, 
\left\{\theta (p^+) \, \left[V (x_t) - 1\right] -
\theta (- p^+) \, \left[V^\dagger (x_t) - 1\right]
\right\}
\ee
the quark propagator is then given by 
\be
\label{eq:prop-gen}
S_F (p,q) = (2 \pi)^4 \delta^4 (p - q)\, S_F^0 (p) +  
S_F^0 (p)\, \tau_F (p,q) \, S_F^0 (q)
\ee
where we have added the free propagator for completeness
and the scattering amplitude is related to $\tau_F$ via 
\be
\label{eq:amp-prop}
i \mathcal{M} (p,q) = \ubar (q)\, \tau_F (p,q)\, u (p)\, .
\ee

\section{Multiple scattering at small and large $x$}

It should be clear from the derivation in the last section 
that the target fields denoted by $S$ have only small $P^+$ modes. 
Furthermore, since we are considering scattering from the 
small $x$ fields of the target, they carry small fraction of 
the target energy $P^-$ and therefore can not impart large $P^-$ 
into the projectile. So the only momentum exchanged is 
transverse momentum which is $\mathcal{O} ({p_t\over p^+})$.
Therefore these soft multiple scatterings  
can not cause a big deflection of the projectile which 
continues to travel along a straight line in $x^+$ direction 
and remains at the same transverse coordinate $x_t$. 

This fact allows one to re-sum 
the soft multiple scatterings into a Wilson line. Physical observables 
in DIS and the hybrid formulation of particle production in 
proton-proton and proton-nucleus collisions involve multi-point 
correlators of these Wilson lines; the most prominent one being 
the dipole scattering cross section which is the trace of two 
Wilson lines in the fundamental representation and satisfies the 
JIMWLK/BK evolution equation. In this sense Wilson lines are the 
effective degrees of freedom at small $x$.

Clearly to have any hope of merging the 
full DGLAP physics, i.e. exact splitting functions, with the JIMWLK/BK 
evolution which takes the small $x$ limit of the splitting functions, 
one needs to relax the small $x$ approximation and include scattering 
from the large 
$x$ modes of the target. One would also need to figure out what the 
effective degrees of freedom are in this more general kinematics where
large $x$ modes of the target are included. This is our main goal in 
this paper. To this end we propose to go beyond scattering from small 
$x$ fields by including one general field which can carry any momentum 
fractions $x$. Such a general field will clearly depend on all 
$4$-coordinates and all of its components (subject to the gauge condition) 
will participate in the scattering. Therefore we consider scattering of 
a quark on a target when in addition to the small $x$ fields one "all $x$" 
field participates. Since this general field will carry large
(as well as small) $P^-$, it can impart a large $p^-$ momentum to the 
projectile and cause a large angle deflection so that the scattered quark 
may have parametrically large transverse momentum 
($p^- = {p_t^2 \over 2 p^+}$) and lose some (or even all) of its $p^+$ 
momentum (rapidity). We expect 
that we may not have to include more than one scattering from an all $x$ 
field and that higher number of such scatterings would be suppressed by 
the coupling constant. In this paper we will ignore higher number of 
scatterings from the all $x$ field but intend to come back to this 
point in the future as it may be relevant for gauge invariance at large $x$. 

\subsection{A hard scattering after multiple soft scatterings}
   
We therefore consider scattering of a quark from an all $x$ field
which we denote $A^\mu = A^\mu (x^+, x^-, x_t)$, the single scattering
amplitude is shown in Fig. (\ref{fig:1hard-scatt}). From now on we 
will refer to the all $x$ field as hard and the small $x$ fields 
as soft so that hard or soft refers to momentum fraction $x$ carried
by a field.  

\begin{figure}[h]
  \centering
  \includegraphics[width=.5\textwidth]{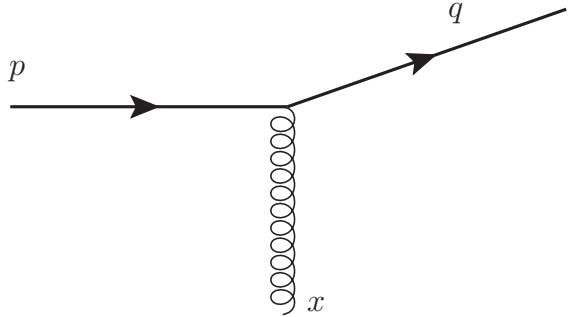}
  \caption{\it One hard  scattering of a quark.}
  \label{fig:1hard-scatt}
\end{figure}

The result is
\bea
i \mathcal{M}_1 &=& (i g) \int d^4 x\, e^{i (q - p) x}\, \ubar (q)\, 
\left[ \slA (x)\right]\, u(p) \nn
&=&
(i g) \int d^4 x\, e^{i (q - p) x}\, \ubar (q)\, 
\left[ \slA (x)\, {\slp \over 2 p^+}\, \sln\right]\, u(p)
\eea
where we have used the Dirac equation to arrive at the second line.
We can now successively include more scatterings from the small $x$ 
fields $S$. Let us consider the case that all such soft scatterings 
happen before the hard scattering. Including one such soft scattering 
gives
\be
i \mathcal{M}_2 = (i g)^2 \int d^4 x\, d^4 x_1 \, \int {d^4 p_1 \over (2 \pi)^4}\, 
e^{i (p_1 - p) x_1}\, 
e^{i (q - p_1) x}\, 
\ubar (q)\, \left[ \slA (x) \, {i \slpone \over p_1^2 + i \epsilon}\, \sln \, S (x_1) 
\right]\, u(p)
\ee
shown in Fig. (\ref{fig:1hard-1soft-scatt}).
\begin{figure}[h]
  \centering
  \includegraphics[width=.5\textwidth]{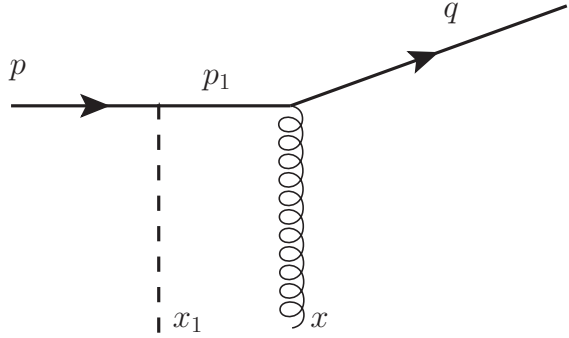}
  \caption{\it One soft scattering before a hard one.}
  \label{fig:1hard-1soft-scatt}
\end{figure}

As before we can perform the $x_1^-$ integration since the soft field $S$ 
does not depend on it, and use the resulting delta function to set 
$p_1^+ = p^+$. Next integration over $p_1^-$ can be done using contour 
integration which gives
\be
( - i) \, \theta (x^+ - x_1^+) \, {1 \over 2 p^+}\, 
e^{i {p_{1t}^2 \over 2 p^+} (x_1^+ - x^+)}\, .
\ee
We get
\bea
i \mathcal{M}_2 &=&  (i g)^2 \int d^4 x\, d^2 x_{1t} \, d x_1^+ \, 
\theta (x^+ - x_1^+)\,
\int {d^2 p_{1t} \over (2 \pi)^2} \, 
e^{i (q - p_1) x} \, e^{- i (p_{1t} - p_t) \cdot x_{1t}}
 \nn
&&
\ubar (q)\, \left[ \slA (x) \, {\slpone \over 2 p^+}\, \sln \, S (x_1^+, x_{1t}) 
\right]\, u(p)
\eea
where $p_1^- = {p_{1t}^2 \over p^+}$ and $p_1^+ = p^+$. This procedure can now be repeated
to include any number of soft scatterings which happen before the hard one. For example, 
inserting two soft scatterings before the hard one is shown in Fig. 
(\ref{fig:1hard-2soft-scatt}) 
\begin{figure}[h]
  \centering
  \includegraphics[width=.5\textwidth]{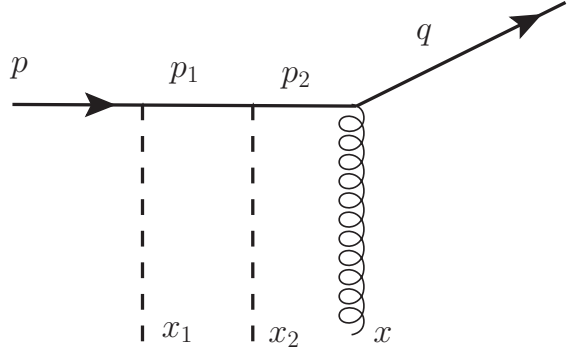}
  \caption{\it Two soft scatterings before a hard one.}
  \label{fig:1hard-2soft-scatt}
\end{figure}
and gives
\bea
i \mathcal{M}_3 &=&  (i g)^3 \!\!\! \int d^4 x\, d^2 x_{1t} \, d x_1^+ \, d x_2^+ \, 
\theta (x^+ - x_2^+) \, \theta (x_2^+ - x_1^+)\! 
\int {d^2 p_{2t} \over (2 \pi)^2} \, e^{i (q - p_2) x} \, 
e^{- i (p_{2t} - p_{1t}) \cdot x_{1t}}
\nn
&&
\ubar (q)\, \left[ \slA (x) \, {\slptwo \over 2 p_2^+}\, \sln \, S (x_2^+, x_{1t})\,
{\slpone \over 2 p_1^+}\, \sln \, S (x_1^+, x_{1t})\,
\right]\, u(p)
\eea
where the soft fields are now at the same transverse coordinate as before. 
It should be now clear how to repeat this procedure by adding more and more
soft scatterings. The amplitude for $n$ soft scatterings followed by a 
hard one is shown in Fig. (\ref{fig:1hard-nsoft-scatt}),  
\begin{figure}[b]
  \centering
  \includegraphics[width=.7\textwidth]{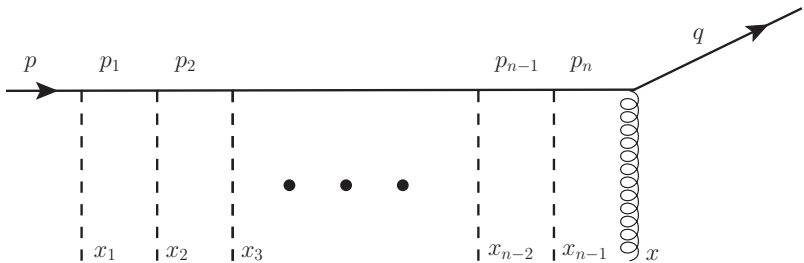}
  \caption{\it N soft scatterings before a hard one.}
  \label{fig:1hard-nsoft-scatt}
\end{figure}
After re-summing all the contributions from $n$ soft
scatterings followed by a hard one we get
\be
i \mathcal{M} =  (i g) \int d^4 x\, d^2 z_t \, 
\int {d^2 k_t \over (2 \pi)^2} \, e^{i (q - k) x} \, 
e^{- i (k_t - p_t) \cdot z_t}
\ubar (q)\, \left[ \slA (x) \, {\slk \over 2 k^+}\, \sln \, V (z_t, x^+) 
\right]\, u(p)\, .
\label{eq:1hard-after-nsoft}
\ee
where $k^+ = p^+, k^- \equiv {k_t^2 \over k^+}$ (momentum $k$ is on shell) 
and the Wilson line now extends to $x^+$,
\be
V (z_t, x^+) \equiv \hat{P}\, 
\exp \left\{i g \int_{- \infty}^{x^+} d z^+ \, S^-_a (z_t, z^+)\, t_a\right\}
\ee
and $S^- (z_t, z^+) = n^-\, S (z_t, z^+)$ as before. 

\noindent This has a simple interpretation; the quark 
propagates through the soft
field and scatters eikonally from the target starting from 
$z^+ = - \infty$ until the point $x^+$ where the hard scattering takes place.  
As before multiple soft scatterings do not change the transverse coordinate of 
the projectile quark and re-sum into a Wilson line.
 
\subsection{Multiple soft scatterings after a hard one} 
In the eikonal approximation used for the soft multiple scatterings the
final state quark can gain some transverse momentum but not much as 
$p_t \ll p^+$. Furthermore its $p^+$ is conserved and does not change which
also means it has negligibly small $p^-$. However, a hard scattering such
as the one considered above can in principle impart large $p^-$ and $p_t$ to the
projectile (and $P^+, p_t$ to the target) so that after the hard scattering
the quark will not necessarily propagate along its original trajectory but
can be deflected by a large angle and lose some or even all of its $p^+$. This
means we must take into account the possibility that the scattered quark is 
not moving along the original (positive) z direction but can now move in an
arbitrary direction with projections on all $x, y, z$ coordinates. Equivalently
the scattered quark will have momenta $p^+, p^-, p_t$ without any restriction
on their magnitude (still subject to being on shell). For example, in the extreme
case of back scattering $p^+ \rightarrow 0$ while $p^-$ will be very large. Or 
in the case of scattering at right angle we will have $p^+ \sim p^- \sim p_t$. 
So it seems that we can not use our eikonal methods after the hard scattering 
any more.

However one can still our eikonal results {\it if we work in the light cone 
frame of the scattered quark} so that light cone frame of the scattered quark is
rotated (in $3$ dimensions) with respect to the light cone frame of the incoming
quark. We define the direction of motion of the scattered quark to be the new 
positive $z$ axis and label it as $\bar{z}$. The direction of propagation of 
the scattered quark $\bar{z}$ will now be related to it $x,y,z$ in the original 
frame (defined by the incoming quark) via the standard rotation matrix 
$\mathcal{O}$ in $3$ dimensions.
\be
\left(\begin{array}{c}
\bar{x} \\
\bar{y} \\
\bar{z}
\end{array} \right)
 = 
\mathcal{O} \,
\left(\begin{array}{c}
x  \\
y \\
z 
\end{array} \right)
\ee
The same rotation matrix will allow us to express the momentum vector 
of the scattered quark in the rotated frame in terms of quantities 
in the original frame. The rotation matrix $\mathcal{O}$ will depend 
on azimuthal and polar angles of the scattered quark with respect to the
original frame which can be related to energy and rapidity of the 
final state quark. 

In the new coordinate system the scattered quark is moving along
positive $\bar{z}$ so that it will have a large $p_{\bar{z}}$ but 
no $p_{\bar{x}}$ or $p_{\bar{y}}$. However it will possibly have 
all momentum components when expressed in terms old frame quantities 
$p_x, p_y, p_z$. One can then define the new light cone coordinates 
$\bar{x}^+, \bar{x}^-, \bar{x}_t$ and light cone momenta 
$\bar{p}^+, \bar{p}^-, \bar{p}_t$ in the new frame in the standard way. 
To facilitate this we define a new light cone vector $\bar{n}$ which 
projects out the plus component in the new frame, so that 
$\bar{n}\cdot \bar{p} = \bar{p}^+$.

The scattered quark now moves along the new plus direction $\bar{x}^+$ and has only
one large momentum component $\bar{p}^+$. Therefore it will couple only to the minus
component of the field expressed in the new coordinates 
$\bar{S}^- = \bar{n}^- \, S(\bar{x}^+, \bar{x}_t)$. It is important to realize that
even though only the $-$ component of the field in the new coordinate system is 
involved, the rotation matrix will generate the other Lorentz components if one 
expresses the soft field $\bar{S}^-$ in terms of the fields in the old coordinate 
system via $\bar{S}^\mu = \Lambda^\mu_{\,\, \nu} \, S^\nu$ where the only non-zero 
components of the Lorentz transformation matrix $\Lambda^\mu_{\,\, \nu}$ are the
elements of the $3$-dimensional rotation sub-group.

To illustrate this we consider the case when there is a hard scattering first, followed
by multiple soft scatterings. The amplitude for one soft scattering after the hard
one, shown in Fig. (\ref{fig:1soft-1hard-scatt}), can be written as
\begin{figure}[h]
  \centering
  \includegraphics[width=.5\textwidth]{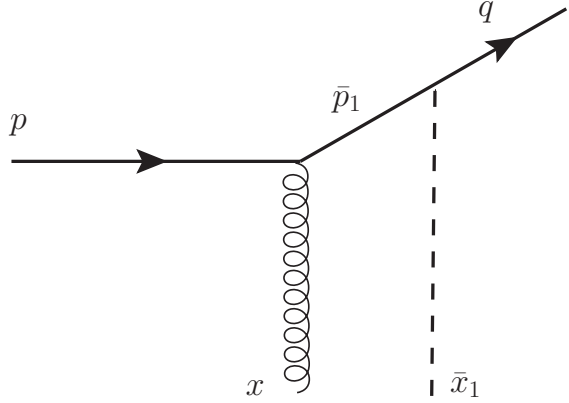}
  \caption{\it One soft scattering after a hard one.}
  \label{fig:1soft-1hard-scatt}
\end{figure}

\be
i \mathcal{M}_2 = (i g)^2 \int d^4 x\, d^4 \bar{x}_1 \, 
\int {d^4 \bar{p}_1 \over (2 \pi)^4}\, 
e^{i (\bar{p}_1 - p) x}\, 
e^{i (\bar{q} - \bar{p}_1) \bar{x}_1}\, 
\ubar (\bar{q})\, \left[ 
\slnbar \, S (\bar{x}_1) \, {i \slpbarone \over \bar{p}_1^2 + i \epsilon}\,
\slA (x) \right]\, u(p)\, .
\ee
All the steps we went through in section (\ref{sec:eikonal}) can now be repeated; 
the field $S (\bar{x}_1)$ is independent of $\bar{x}_1^-$ so that integration over
$\bar{x}_1^-$ can be performed leading to a delta function of 
$2 \pi\, \delta (\bar{p}_1^+ - \bar{q}^+)$ which allows one to perform the 
$\bar{p}_1^+$ integration setting $\bar{p}_1^+ = \bar{q}^+$. One then performs 
the contour integration over $\bar{p}_1^{\,-}$ noticing that it is next to $\slnbar$
which puts $\bar{p}_1$ on shell and gives a factor of $\theta (\bar{x}_1^+ - x^+)$,
 we get
\bea
i \mathcal{M}_2 &=&  (i g)^2 \int d^4 x\, d^2 \bar{x}_{1t} \, d \bar{x}_1^+ 
\, \theta (\bar{x}_1^+ - x^+)\,
\int {d^2 \bar{p}_{1t} \over (2 \pi)^2} \, e^{i (\bar{p}_1 - p) x} \, 
e^{- i (\bar{q}_t - \bar{p}_{1t}) \cdot \bar{x}_{1t}}
 \nn
&&
\ubar (\bar{q})\, \left[ 
\slnbar \, S (\bar{x}_1^+, \bar{x}_{1t}) \, {\slpbarone \over 2 \bar{p}_1^+} \,
\slA (x)
\right]\, u(p)
\eea
with 
$\bar{p}_1^+ = \bar{q}^+$ and $\bar{p}_1^- = {\bar{p}_{1t}^2 \over 2 \bar{p}_1^+}$.
By repeating this procedure for further soft scatterings as shown in 
Fig. (\ref{fig:nsoft-1hard-scatt}),
\begin{figure}[h]
  \centering
  \includegraphics[width=.5\textwidth]{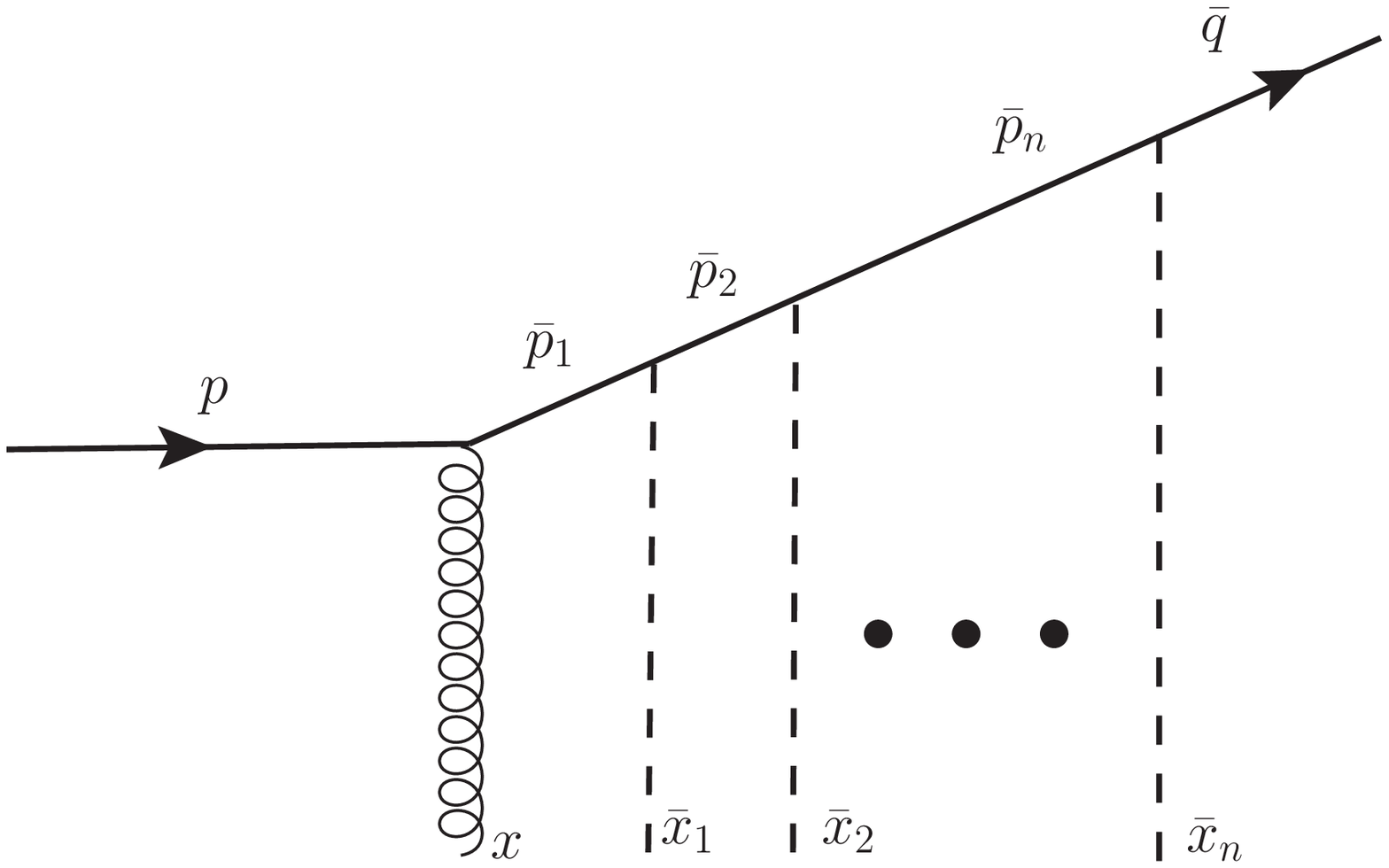}
  \caption{\it N soft scatterings after a hard one.}
  \label{fig:nsoft-1hard-scatt}
\end{figure}
and re-summing them we get
\be
i \mathcal{M} =  (i g) \int d^4 x\, d^2 \bar{z}_t \, 
\int {d^2 \bar{k}_t \over (2 \pi)^2} \, e^{i (\bar{k} - p) x} \, 
e^{- i (\bar{q}_t - \bar{k}_t) \cdot \bar{z}_t}
\ubar (\bar{q})\, \left[ 
\overline{V} (x^+, \bar{z}_t) \, \slnbar \, {\slkbar \over 2 \bar{k}^+} \,
\slA (x)
\right]\, u(p)
\ee
where 
$\bar{k}_1^+ = \bar{q}^+$ and $\bar{k}_1^- = {\bar{k}_t^2 \over 2 \bar{k}_1^+}$
and the Wilson line is now 
\be
\overline{V} (x^+, \bar{z}_t) \equiv \hat{P}\, 
\exp \left\{i g \int_{x^+}^{+\infty} d \bar{z}^+ \, S^-_a 
(\bar{z}_t, \bar{z}^+)\, t_a\right\}
\ee

Again this expression has a simple interpretation; the projectile quark 
undergoes a hard scattering which changes its direction (possibly in all 
$3$ dimensions) after which it undergoes multiple soft scatterings when 
expressed in terms of coordinates in the new frame.

\subsection{Multiple soft scatterings before and after a hard scattering}
We now have all the ingredients to write the general expression for 
scattering of a projectile quark from the target including soft (small $x$)
scatterings to all order and hard (all $x$) to the first order. This is shown
in Fig. (\ref{fig:nsoft-1hard-nsoft-scatt}) where the final state quark may now
have a large $q^-$ or $q_t$ as measured in the incoming quark's frame (but has 
only a large momentum in $\bar{z}^+$ direction in the rotated frame).
\begin{figure}[h]
  \centering
  \includegraphics[width=1.0\textwidth]{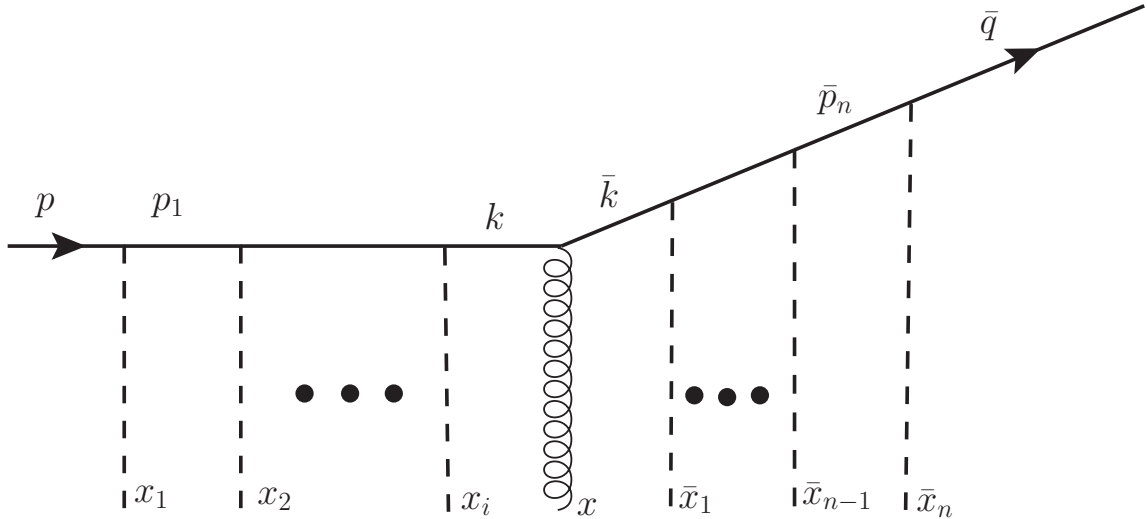}
  \caption{\it N soft scatterings before and after a hard one at $x$.}
  \label{fig:nsoft-1hard-nsoft-scatt}
\end{figure}

The procedure
is the same as above, one considers multiple scatterings from the soft fields
which are independent of the $"-"$ coordinate leading to conservation of the 
$"+"$
momentum across a soft scattering, and then doing the contour integration 
over the $"-"$ component of the intermediate momentum which puts the quark
line on shell. All the terms of the form ${p_t \over p^+}\, x^+$ or 
${\bar{p}_t \over \bar{p}^+}\, \bar{x}^+$ are neglected which allows one to 
do the transverse coordinate integration over the soft fields. The only exception
is if there is a "mis-match" of barred and un-barred momenta and coordinates
in the exponentials which are then kept, for example exponents like 
$ \bar{p}^-\, x^+$ may have large components when expressed fully in terms 
of quantities in the old frame. The rest of details are the same as before;
during soft multiple scatterings the transverse coordinate of the quark 
remains the same and
there is ordering in the direction of $x^+$ or $\bar{x}^+$ which allows
one to re-sum the multiple scatterings into a Wilson line. As a result we
have $x_{1t} = x_{2t} = \cdots = x_{it}= z_t$ and 
$\bar{x}_{1t} = \bar{x}_{2t} = \cdots = \bar{x}_{nt}= \bar{z}_t$. Therefore 
the full amplitude can be re-summed and written as
\bea
i \mathcal{M} &=&  (i g) \int d^4 x\, d^2 z_t \, d^2 \bar{z}_t \, 
\int {d^2 k_t \over (2 \pi)^2} \, {d^2 \bar{k}_t \over (2 \pi)^2} \, 
e^{i (\bar{k} - k) x} \, 
e^{- i (\bar{q}_t - \bar{k}_t)\cdot \bar{z}_t}\, 
e^{- i (k_t - p_t)\cdot z_t}
 \nn
&&
\ubar (\bar{q})\, \left[ 
\overline{V} (x^+, \bar{z}_t) \, \slnbar \, {\slkbar \over 2 \bar{k}^+} \,
\slA (x)\, 
{\slk \over 2 k^+} \, \sln \, V (z_t, x^+)  
\right]\, u(p)
\label{eq:nsoft-before-1hard-after-nsoft}
\eea
with $k^+ = p^+, k^- = {k_t^2 \over 2 k^+}$ and 
$\bar{k}^+ = \bar{q}^+, \bar{k}^- = {\bar{k}_t^2 \over 2 \bar{k}^+}$. 

As before the scattering amplitude can be written in terms of the 
interaction part of the propagator via
\be
i \mathcal{M} (p,\bar{q}) = \ubar (\bar{q})\, \tau_F (p,\bar{q})\, u (p)
\ee
in terms of which the full (Feynman) propagator is 
\be
\label{eq:prop-gen-both}
S_F (p,\bar{q}) = (2 \pi)^4 \delta^4 (p - \bar{q})\, S_F^0 (p) +  
S_F^0 (p)\, \tau_{hard} (p,\bar{q}) \, S_F^0 (\bar{q})
\ee
with interacting part of the propagator is then given by 
(including the contribution where quark is moving backward 
in $x^+$ which is needed for the Feynman propagator)
\bea
\label{eq:tau_{hard}}
\tau_{hard} (p,\bar{q}) &\equiv&  (i g) \int d^4 x\, 
\int {d^2 k_t \over (2 \pi)^2} \, {d^2 \bar{k}_t \over (2 \pi)^2} \, 
d^2 z_t \, d^2 \bar{z}_t \, e^{i (\bar{k} - k) x} \, 
e^{- i (\bar{q}_t - \bar{k}_t)\cdot \bar{z}_t}\, 
e^{- i (k_t - p_t)\cdot z_t}\, \nn
&&
\bigg\{\theta (p^+) \, \theta (\bar{q}^+) \, V (z_t, x^+)\,
\sln \, {\slk \over 2 k^+} \, \slA (x) \, {\slkbar \over 2 \bar{k}^+}\,
\slnbar \, \overline{V} (x^+, \bar{z}_t) - \nn
&&
\theta (- p^+) \, \theta (- \bar{q}^+) \, 
 V^\dagger (z_t, x^+)\,
\sln \, {\slk \over 2 k^+} \, \slA (x) \, {\slkbar \over 2 \bar{k}^+}\,
\slnbar \, \overline{V}^\dagger (x^+, \bar{z}_t) 
\bigg\}\, .
\eea 
It is straightforward to check that this expression re-sums all the 
diagrams shown in Fig. (\ref{fig:expanded-scatt}) by expanding 
the Wilson lines. The dashed lines correspond to scatterings 
from the soft fields carrying small $x$ fractions of the target 
energy whereas the wavy line corresponds to scattering from a hard field 
carrying arbitrary $x$ fraction of the target energy. It describes 
propagation of an energetic quark along longitudinal direction $z^+$ and 
undergoing multiple soft interactions described by $ V (z_t, x^+)$ until 
the location $x^\mu$ at which point it undergoes a potentially hard 
(large $x$) scattering which changes it direction. After the hard 
scattering it propagates again along the new longitudinal direction 
$\bar{z}^+$ and undergoing multiple scatterings described by 
$\overline{V} (x^+, \bar{z}_t)$. The location of the hard scattering 
$x^\mu$ is integrated from $- \infty$ to $+ \infty$ (or from $- R$ to
$R$ where $2 R$ is the diameter of the nucleus/size of the medium) 
so that the hard scattering can happen at any $x^+$, at the front, 
back or middle of the nucleus/medium, followed or preceded by multiple 
soft scatterings. 

\begin{figure}[h]
  \centering
  \includegraphics[width=1.0\textwidth]{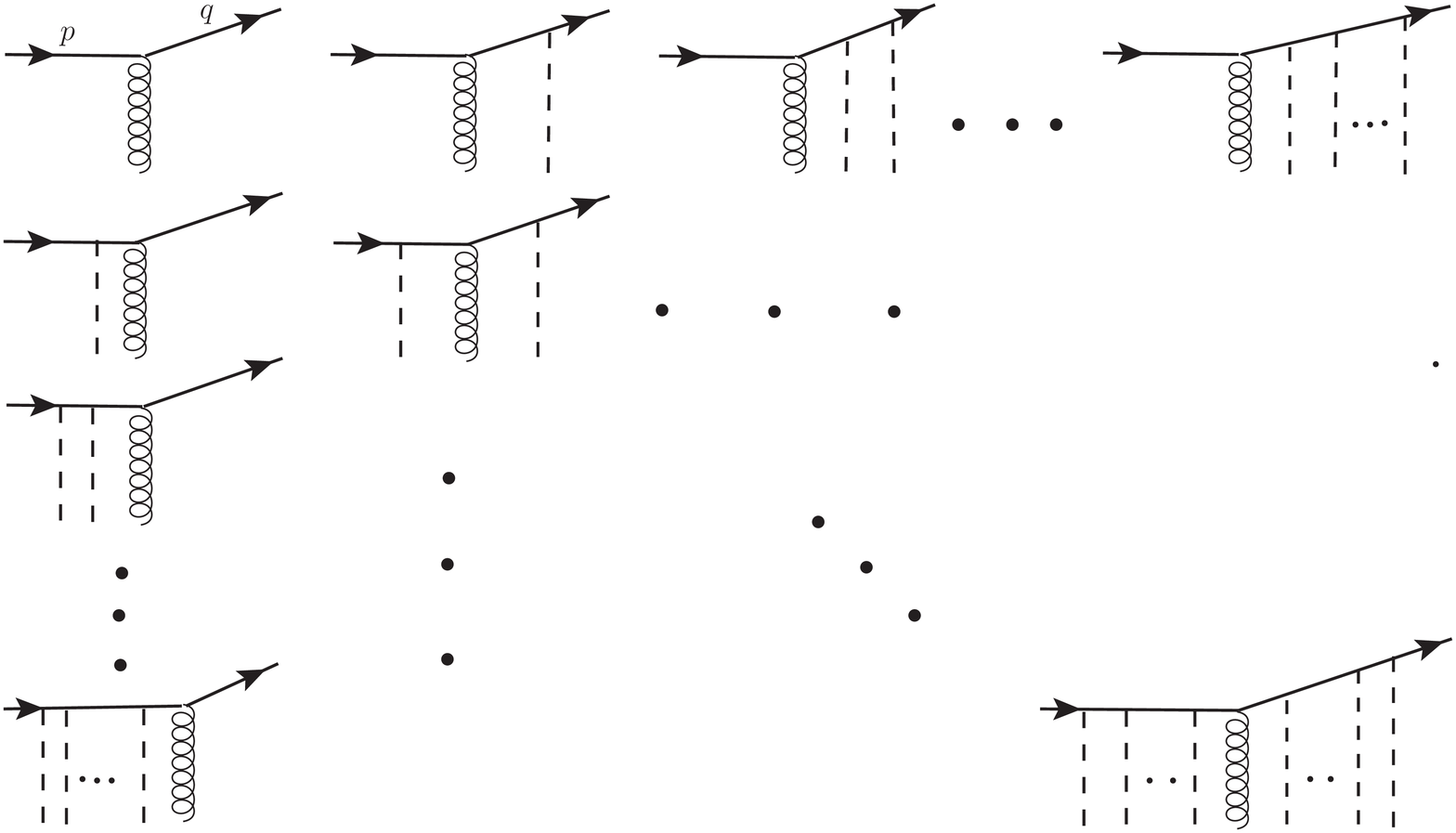}
  \caption{\it Diagrams re-summed by eq. (\ref{eq:tau_{hard}}).}
  \label{fig:expanded-scatt}
\end{figure}

There is one more technical but essential point that we need to address; 
the coupling 
between the quark and gluon fields we have considered is of the form
$\bar{\Psi}\, \slA\, \Psi$ which is not gauge invariant. Instead one
needs to use the full covariant coupling $\bar{\Psi}\, \slD\, \Psi$
in the gauge invariant Lagrangian.
This does not matter in the eikonal limit since replacing any of the 
soft fields by the normal derivative gives a vanishingly small correction. 
However in our more general case it can act on the Wilson line in the 
rotated frame and give
contributions which are of the same order as considered here. Therefore
to have the interacting part of the effective propagator transform 
covariantly one needs to include it. To do so is trivial and amounts 
to replacing the hard field $\slA$ in our step by step derivation by 
the covariant derivative $\slD$. Therefore our expression in eq. 
(\ref{eq:tau_{hard}}) is modified to
\bea
\label{eq:tau_{cov}}
\tau_{hard} (p,\bar{q}) &\equiv& \int d^4 x\, 
\int {d^2 k_t \over (2 \pi)^2} \, {d^2 \bar{k}_t \over (2 \pi)^2} \, 
d^2 z_t \, d^2 \bar{z}_t \, e^{i (\bar{k} - k) x} \, 
e^{- i (\bar{q}_t - \bar{k}_t)\cdot \bar{z}_t}\, 
e^{- i (k_t - p_t)\cdot z_t}\, \nn
&&
\bigg\{\theta (p^+) \, \theta (\bar{q}^+) \, V (z_t, x^+)\,
\sln \, {\slk \over 2 k^+} \, \slD_x \, {\slkbar \over 2 \bar{k}^+}\,
\slnbar \, \overline{V} (x^+, \bar{z}_t) - \nn
&&
\theta (- p^+) \, \theta (- \bar{q}^+) \, 
 V^\dagger (z_t, x^+)\,
\sln \, {\slk \over 2 k^+} \, \slD_x \, {\slkbar \over 2 \bar{k}^+}\,
\slnbar \, \overline{V}^\dagger (x^+, \bar{z}_t) 
\bigg\}\, .
\eea    
This is our final result for the interacting part of the quark propagator.
This expression has the interesting property that it includes both 
partially and fully coherent multiple scatterings of the quark 
from the target and as such generalizes the fully coherent multiple 
scattering in eikonal approximation. 

It is very interesting to consider the soft limit of this
expression, i.e. the limit when the all $x$ field $\slA$ carries 
a small $x$ fraction of the target energy. To do so we need to replace 
$\slA (x^+, x^-, x_t)\rightarrow \sln\, S(x^+, x_t)$ and take all 
barred quantities to be un-barred, for example, 
$\slnbar \rightarrow \sln$ and $\bar{k} \rightarrow k$. It is easy then
to show eq. (\ref{eq:tau_{cov}}) vanishes in this limit due to action of
the covariant derivative on a Wilson lines under a parallel transport
along the direction of motion ($n\cdot D_x\, V (x^+, x_t) = 0$). 
This property of eq. (\ref{eq:tau_{cov}}) suggests that to have a quark 
propagator which can be used for construction of physical quantities
such as structure functions $F_2$ and $F_L$ valid for any $x$ one should 
include both contributions from purely eikonal scattering as given by eq. 
(\ref{eq:tau_eik}) and hard scattering given by eq. (\ref{eq:tau_{cov}}). 
Therefore we take the most general interaction part of the effective 
propagator to be
\be
\tau_{all \, x} \equiv \tau_F + \tau_{hard}
\label{eq:tauall}
\ee  
which is shown symbolically in Fig. (\ref{fig:tauallx}) 
\begin{figure}[h]
  \centering
  \includegraphics[width=1.0\textwidth]{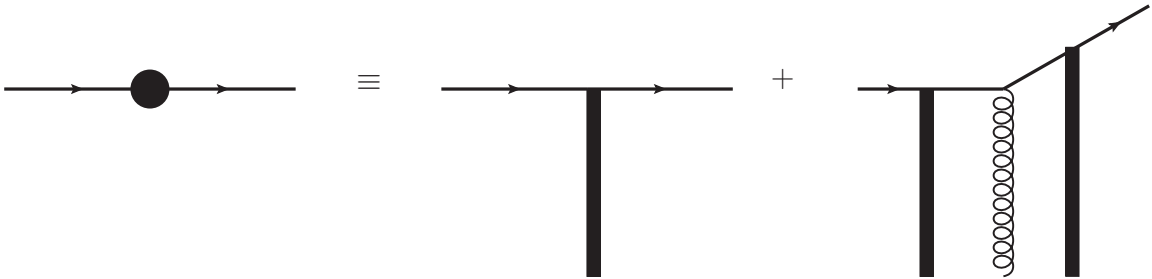}
  \caption{\it Interaction part of the effective propagator in eq. 
(\ref{eq:tauall}).}
  \label{fig:tauallx}
\end{figure}
and where the thick solid lines denote a Wilson line representing multiple
soft scatterings from the target to all orders.

It should be noted that while the projectile quark scatters from both 
low and high $x$ modes of the target color fields, we have not included
interactions between the different $x$ modes. In this sense the small and
large $x$ fields are non-interacting with respect to each other. For 
instance it should be possible for a soft field to couple to a hard field 
as shown in Fig. (\ref{fig:selfinteract-1hard-1soft-scatt})
\begin{figure}[h]
  \centering
  \includegraphics[width=0.5\textwidth]{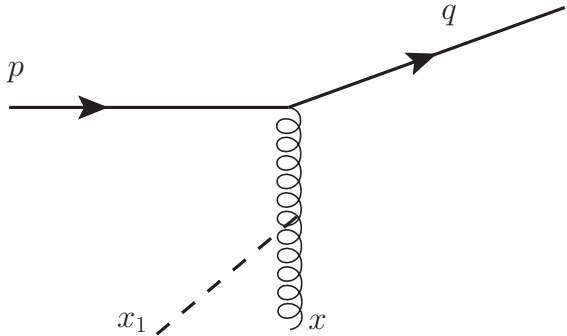}
  \caption{\it Interaction of a soft field with a hard one.}
  \label{fig:selfinteract-1hard-1soft-scatt}
\end{figure}

A 
preliminary study shows that the first coupling of 
the soft field to the hard field forces the two fields to be at the same 
position in coordinate space. Contributions of diagrams like this are 
currently under investigation and will be reported elsewhere.

\section{Discussion and summary}

We have constructed an effective quark propagator in the background
of color fields which can carry both small and large $x$ energy 
fractions of the target from which the quark scatters. This generalizes 
the known results for the quark propagator in the background of 
a color field carrying small $x$ fraction of the target energy based
on eikonal approximation. 

The constructed effective propagator can be useful in many ways; first,
it includes the possibility that a projectile quark can scatter from the 
more energetic (large $x$) partons of the target and get deflected by 
a large angle (high $p_t$). This may already give significant contributions 
to single inclusive particle production in high energy collisions~\cite{jjm-dA} 
when one uses the hybrid 
formalism. It may also lead to significant final state momentum anisotropy
at high $p_t$ if one couples it to realistic nuclear geometries that 
include fluctuations in the target nucleus/medium~\cite{bsrv}.

Furthermore, one can use this propagator to calculate the QCD structure 
functions $F_2$ and $F_L$ following the procedure suggested in \cite{mv-f2}.
This would generalize the standard dipole picture of DIS at small $x$ by
including the contributions of the large $x$ gluons in the target. This work
is in progress and will be reported elsewhere~\cite{jjm-f2}.

To proceed further, one can use this method to calculate the effective 
gluon propagator in a similar fashion. One can then generalize the 
hybrid formalism for particle production in asymmetric high energy 
proton-proton and proton-nucleus collisions and extend the validity 
of the hybrid formalism to high $p_t$, keeping in mind that high $p_t$ in
this context is equivalent to large $x$. Hence it would include 
contributions of both fully and partially coherent multiple scatterings.

It will be interesting to see if one can use our results to generalize 
and extend the McLerran-Venugopalan (small $x$) effective 
action~\cite{mv,jv} to include contributions of large $x$ gluons. 
If so one would then be able to apply the MV model to not only low~\cite{kmw}
but also high $p_t$ particle production in high energy heavy ion collisions. 
This would allow us to 
investigate and quantify the contributions of early time dynamics to 
jet energy loss in Quark-Gluon Plasma formed in high energy heavy ion 
collisions.

Perhaps most significantly, we hope the physical quantities computed  
using this effective propagator will serve as the Leading Order 
expressions which can then be used for deriving Leading Log evolution 
equations which would have DGLAP and JIMWLK evolution equations in 
appropriate limits~\cite{jjm-f2}.

\section*{Acknowledgments}
We acknowledge support by the DOE Office of Nuclear Physics
through Grant No.\ DE-FG02-09ER41620 and by the Idex Paris-Saclay 
though a Jean d'Alembert grant. We would like to thank N. Armesto, 
F. Gelis, E. Iancu, C. Lorc\'e, C. Marquet, A.H. Mueller, S. Munier, 
B. Pire, C. Salgado, G. Soyez and B. Xiao for critical questions, 
illuminating discussions and helpful suggestions.

\end{document}